\documentclass[useAMS,usenatbib]{mn2e}
\usepackage{graphicx,epsfig,psfig}
\usepackage{url}
\title[RAMSES-CH: A New Chemodynamical Code for Cosmological Simulations]{RAMSES-CH: A New Chemodynamical Code for Cosmological Simulations}
\author[Few et~al.]{C.G. Few$^{1}$\thanks{E-mail:
cgfew@uclan.ac.uk}, S. Courty$^{2}$, B.K. Gibson$^{1,3,4}$, D. Kawata$^{5}$, 
F. Calura$^{1,6}$ and R. Teyssier$^{7,8}$\\
$^{1}$Jeremiah Horrocks Insitute, University of Central Lancashire, 
Preston, PR1~2HE, UK\\
$^{2}$Universit\'e de Lyon; Universit\'e Lyon~1, Observatoire de Lyon, 9 avenue Charles Andr\'e, Saint-Genis Laval, F-69230, France;\\
 CNRS, UMR 5574, Centre de Recherche Astrophysique de Lyon; Ecole Normale Sup\'erieure de Lyon\\
$^{3}$Monash Centre for Astrophysics, School of Mathematical Sciences, 
Monash University, Clayton, VIC, 3800, Australia\\
$^{4}$Department of Astronomy \& Physics, Saint Mary's University, Halifax,
Nova Scotia, B3H~3C3, Canada\\
$^{5}$Mullard Space Science Laboratory, University College London, Holmbury 
St. Mary, RH5 6NT, United Kingdom\\
$^{6}$Istituo Nazional di Astrofisica, Osservatorio Astronomico di Bologna, 
Via Ranzani 1, I-40127, Bologna, Italy\\
$^{7}$Institute for Theoretical Physics, University of Z\"urich, CH-8057, 
Z\"urich, Switzerland\\
$^{8}$UMR AIM, CEA Saclay, 91191 Gif-sur-Yvette, France}

\begin{document}
\date{Submitted}
\pagerange{\pageref{firstpage}--\pageref{lastpage}} \pubyear{2012}
\maketitle
\label{firstpage}

\begin{abstract}
We present a new chemodynamical code -- \textsc{Ramses-CH} -- for 
use in simulating the self-consistent evolution of chemical and 
hydrodynamical properties of galaxies within a fully cosmological 
framework.  We build upon the adaptive mesh refinement code 
\textsc{Ramses}, which includes a treatment of self-gravity, 
hydrodynamics, star formation, radiative cooling, and supernovae 
feedback, to trace the dominant isotopes of C, N, O, Ne, Mg, Si, and 
Fe. We include the contribution of Type~Ia and II supernovae, in 
addition to low- and intermediate-mass asymptotic giant branch stars, 
relaxing the instantaneous recycling approximation. The new chemical 
evolution modules are highly flexible and portable, lending themselves 
to ready exploration of variations in the underpining stellar and 
nuclear physics. We apply \textsc{Ramses-CH} to 
the cosmological simulation of a typical $L_\star$ galaxy, demonstrating 
the successful recovery of the basic empirical constraints regarding, 
[$\alpha$/Fe]--[Fe/H] and Type~Ia/II supernovae rates.
\end{abstract}


\section{Introduction}

The determination of elemental abundance patterns is one of the primary 
diagnostics of galaxy formation, with numerous spatial and temporal 
trends between age, kinematics, and chemistry guiding our insights into 
the underpining physical processes. Observations of abundance ratios 
corroborate our understanding of the nuclear physics governing 
$\alpha$-element production, in that they are produced on shorter 
timescales than iron-peak elements \citep[e.g.][]{carbon87, 
edvardsson93, reddy06, ramirez07}, as a consequence of the 
mass-dependent nuclear burning processes acting within the relevant 
progenitor stars. Galactic chemical evolution (CE) models are predicated upon 
a coupling of these elemental production sites/timescales with 
phenomenological (yet, empirically constrained) parameterisations of 
star formation and gas inflows/outflows.  The resulting predicted 
abundance patterns can be compared directly with observations in order 
to shed light on the formation and evolution of the system under study.

The formalism associated with the semi-numerical approach to galactic 
CE \citep[e.g.][]{talbot71, pagel75, tinsley80, 
matteucci89, carigi94, gibson97, chiappini97, ramirez07} is 
a powerful tool when applied to sub-grid CE 
treatments within fully hydrodynamical simulations. The inclusion of 
CE schemes has been achieved in a number of cosmological 
hydrodynamical codes \citep[e.g.][]{lia02, valdarnini03, kawata03, 
kobayashi04, tornatore04, romeo05, martinezserrano08, oppenheimer08, 
wiersma09, shen10}, each of which are based 
upon smoothed particle hydrodynamics (SPH).
Key lessons can be learned from an examination of the role that CE plays 
in the physics of the interstellar medium. This is manifest in 
the metallicity-dependent radiative cooling rates of plasmas and their 
impact on the efficiency of metal transport throughout the disk and its 
consequent impact on stellar chemo-dynamics \citep{scannapieco05}. This 
impact upon turbulence-driven metal transport can be problematic, in 
light of known issues concerning the ability of conventional 
treatments of SPH to resolve the associated instabilities 
in certain regimes; such problems are ameliorated (though not entirely) by Eulerian 
approaches to fluid dynamics, including adaptive mesh 
refinement (AMR) schemes \citep[e.g.][]{oshea05, agertz07, tasker08}.

In its simplest form, interparticle `mixing' of SPH particles does
not occur, i.e. metal-rich and metal-poor gas particles may 
co-exist near each other without sharing/mixing of their associated
metals.  The impact of this lack of mixing is readily apparent in
a galaxy simulation's metallicity distribution function and
age-metallicity relation, as well as the abundance ratio plane \citep[e.g.][]{pilkington12c}. The 
inclusion of turbulent mixing models within SPH remedies this lack of 
implicit diffusion \citep{shen10}, even if the associated diffusion 
coefficient is a necessary additional free parameter (albeit, informed by turbulence theory).

With the intention of providing a complementary (AMR) approach to extant 
(SPH) chemodynamical and semi-numerical CE models, we 
present what is, to our knowledge, the first cosmological AMR code which 
implements a temporally-resolved feedback and CE 
prescription. Written as a patch to the gravitational cosmological 
N-body and hydrodynamical code \textsc{Ramses}, we now include the 
effects of Type~II supernovae (SNeII), Type~Ia supernovae (SNeIa), and 
low- to -intermediate-mass asymptotic giant branch (AGB) stars, both from 
an energetic and chemical perspective. Nucleosynthetic processes are 
accounted for as a function of progenitor mass and metallicity. 
The CE module is described in \S\ref{model}.  The mechanics of marrying of this 
module to the AMR code \textsc{Ramses}, leading to the self-consistent 
chemodynamical code \textsc{Ramses-CH}, is outlined in \S\ref{ramsesCH}. 
Finally, in \S\ref{fiducial}, we present a demonstration of the ability 
of the code to reproduce basic observational constraints. We emphasise 
that this \emph{Letter} is primarily a methodological description for coupling 
CE with AMR, and that future papers in this series will 
explore the response of \textsc{Ramses-CH} to the various input parameters 
and assembly histories.

\section{Chemical Evolution Model}
\label{model}

\begin{figure}
\includegraphics[width=84mm]{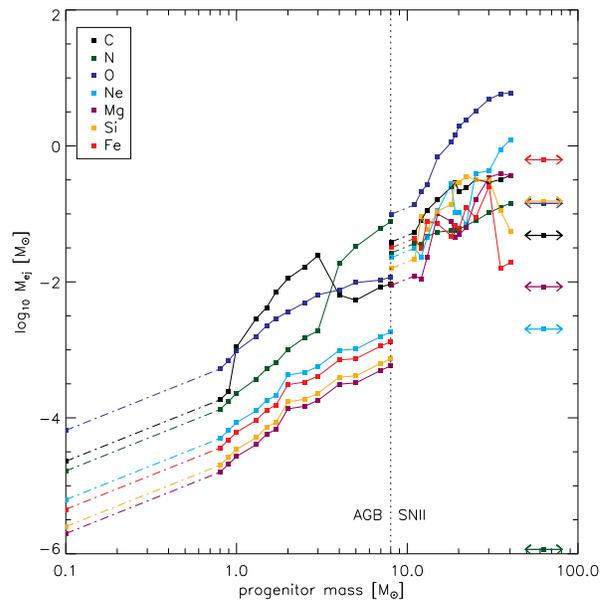}
\caption{Mass of elements ejected by stars as a function of initial 
mass. Also shown are the abundances for a single SNeIa (horizontal arrows) for comparison (the 
position along the abscissa is arbitrary, chosen to avoid conflict 
with other data). The mass above which stars are considered to be 
SNeII progenitors is indicated at 8~M$_\odot$. Data for AGB stars are 
taken from \citet{vdhoek97}, SNeII from \citet{ww95}, and SNeIa from 
\citet{iwamoto99}. Points connected by solid lines denote the original 
data, those connected by dot-dash lines show adopted extrapolations 
to lower masses. Extrapolations are linear and scaled to the mass of the progenitor star.}
\label{ejecta}
\end{figure}

The underlying CE model used to determine the relative 
rates of SNeII:SNeIa:AGB, and the associated chemical enrichment, for a 
stellar population governed by a given IMF, is generated prior to the 
simulation being run.  The resulting look-up tables provide the SNeII, 
SNeIa, and isotopic return rates as a function of time for a range of 
Simple Stellar Population (SSP) metallicities. The code (provided as 
part of the \textsc{Ramses-CH} patch) is flexible, allowing the user to 
readily modify relevant stellar physics, via the importation of 
different SNe and AGB yields, as well as the IMF.

For this first work, we have adopted a fairly 
standard/conservative CE model, employing a 
\citet{kroupa01} IMF with stellar mass limits of 0.1 and 
100~M$_\odot$, respectively.  We also used a SNIa delayed time 
distribution formalism similar to that presented in \citet{kobayashi00} 
and \citet{kawata03}, with the simplification that the IMF slope for 
both the primaries and secondaries was taken to be identical. Stellar 
lifetimes were taken from \citet{kodama97} and are dependent upon both 
mass and metallicity. The yields of SNeII progenitors 
(11$-$40~M$_\odot$), for metallicities spanning Population~III to solar, 
are from \citet{ww95}; for $m$$>$30~M$_\odot$, we adopt the yields 
associated with the Model~B explosion energies, after \citet{timmes95} 
and \citet{kawata03}.  Massive stars in the range 8$-$40~M$_\odot$ are 
assumed to explode as SNeII, where the yields in the range 
8$-$11~M$_\odot$ are found by scaling the elemental mass fractions of the 
11M$_\odot$ stars with the progenitor mass. The same process is used to calculate 
the yields of stars down to 0.1M$_\odot$ using the lowest mass (0.8M$_\odot$) AGB star available in 
\cite{vdhoek97}. The yields adopted are illustrated in Fig.~\ref{ejecta}.

As noted earlier, nucleosynthetic yields for SNeIa were taken from 
\citet{iwamoto99}. In addition to the time constraints imposed by the 
mass range of the secondaries in SNeIa (binary) progenitors, 
\citet{hachisu99} also suggest the use of a metallicity `floor' which 
suppresses the formation of low-metallicity ([Fe/H]$<$$-$1.1) SNIa 
progenitors. In light of the ongoing controversy regarding this putative 
metallicity floor, we have adopted the conventional assumption that 
low-metallicity binaries are capable of forming SNeIa progenitors. These 
SNeIa yields are also noted in Fig.~\ref{ejecta} by the horizontal 
arrows.

\begin{figure}
\includegraphics[width=84mm]{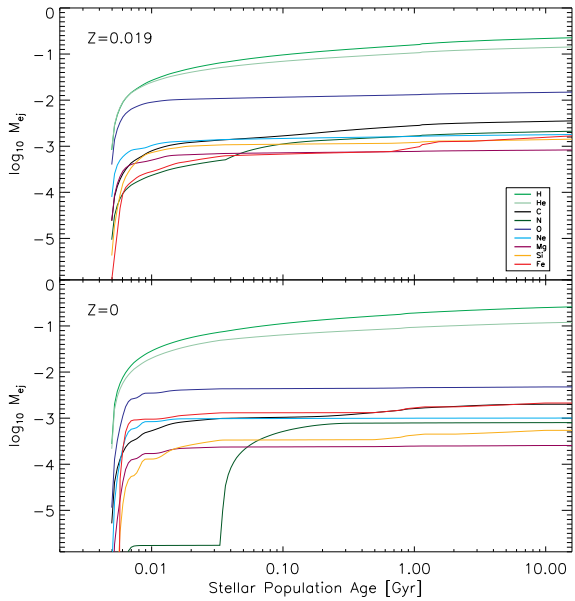}
\caption{Ejection rate of dominant elemental isotopes, per unit stellar 
mass, as a function of age for a \citet{kroupa01} IMF. The upper and lower panels correspond to solar 
and Population~III metallicity simple stellar populations respectively.}
\label{yield_curve}
\end{figure}

The time evolution of the isotopic ejection rate (per unit mass) from an 
SSP is shown in Fig.~\ref{yield_curve}.  The \textsc{Ramses-CH} chemical evolution 
model generates a family of such ejection rates for any combination of 
yield compilation and initial mass function. We have chosen to 
simply show the impact of the choice of one conservative combination of 
parameters.  This should not be 
construed as implying that this combination is necessarily the best 
or unique pairing; it is simply chosen to demonstrate the efficacy 
of the methodology.

\section{Ramses-CH}
\label{ramsesCH}

We have introduced the CE prescription into the v3.07
public release version of \textsc{Ramses} \citep{teyssier02}. Prior to
these enhancements, \textsc{Ramses} tracked the total gas metallicity
(Z), under the assumption of the instantaneous recycling approximation
and treating Z as a passive scalar advected by the hydrodynamical
flow. Our new chemodynamical version (\textsc{Ramses-CH}) also employs
passive scalars in the tracking of the dominant isotopes of H, C, N,
O, Mg, Ne, Si, and Fe, and the chemical composition of the gas from
which the stellar particles form is `tagged' onto the new particles.
As described by \citet{dubois08a}, star particles are created in the
high-density gas ($\rho_{gas}>\rho_{th}$) and spawned by a random
Poisson process following a rate given by $\dot \rho_* =
\epsilon_*\rho_{gas}/t_{ff}$, where $t_{ff}$ is the local free-fall
time of the gas, $(3\pi/32G\rho_{gas})^{1/2}$, and $\epsilon_*$ the
star formation efficiency. Each stellar particle enriches the
surrounding interstellar medium (ISM) according to its individual
chemical history recorded in the look-up tables described in
\S\ref{model}, depending upon the particle's initial mass and
metallicity. Chemical enrichment and feedback processes are treated
simultaneously through a kinetic feedback mode. While we could apply
kinetic feedback to all the stellar populations, we do so only
to those stellar populations that include SNeII events. A
thermal feedback mode is used when the stellar population ages
and enters an AGB and SNIa ``phases''. The kinetic feedback mode
aims at reproducing those expanding gas flows generated by the
collective explosions of massive stars. At each time-step, density,
momentum, energy, and metals are deposited into all gas cells situated
within a feedback-`sphere' of a user-specified radius centered upon
the young star particles. We set up our SNeII feedback-sphere radius
to 2 grid cells. The velocity of the entrained gas, linearly
interpolated with the radius, is given by the number of SNeII events,
the energy generated by each event $\epsilon_{SN} E_{SN}$
($\epsilon_{SN}$ being the efficiency with which the energy
$E_{SN}=10^{51}$~erg couples to the surrounding ISM), and the total
amount of gas to be entrained by the bubble. The entrained gas
includes the gas released/ejected during the SNeII events as well as
the gas swept up by the bubble (Dubois $\&$ Devriendt, private
comm.). This latter component is parametrized as $f_w$ times the
ejected gas mass ($f_w=10$ in this work, corresponding to a mass
loading factor of $\eta_w=1$ for a standard run with a massive star
fraction of $10\%$). The galactic outflow scheme just described is not
applied for the SNIa events and when the stellar population enters an
AGB and/or SNIa phase, feedback processes and chemical enrichment are
handled `locally' (thermally) and confined to the gas cell within which the
stellar particle sits. The feedback algorithm is simplified by 
this approximation; future developments of the code will include kinetic 
SNIa feedback.

This new chemodynamical version of \textsc{Ramses} was then applied in
the generation of a multi-resolved, cosmological simulation of a
late-type disk galaxy (henceforth called, \texttt{109-CH}) with a
virial mass of $7.1\times 10^{11}$~M$_\odot$, whose initial conditions
match those outlined in an earlier study \citep{Sanchez-Blazquez09}.
To remind the reader, the box size for this run was
$L$=20~$h^{-1}$~Mpc, with a dark matter particle mass of $6\times
10^6$~M$_\odot$ in the most refined region. The spatial resolution,
kept roughly constant in physical size during the simulation, reaches
$L/2^{lmax}$=436~pc at z=0, with $lmax$=16 levels of refinement. For
this resolution, we choose a star formation density threshold
$\rho_{th}$ corresponding to 0.3~cm$^{-3}$ and an efficiency of
1$\%$. The same SNe energy coupling efficiency of $\epsilon_{SN}=1$ is
assumed for both SNeII and SNeIa. A polytropic equation of state
$T=T_{th}(\rho/\rho_{th})^{\gamma-1}$ is used in the high-density
areas with a temperature threshold of 2900~K and a polytropic index
$\gamma$=2, allowing the Jeans length to be resolved by more than 4
cells at all times. Note that our feedback parameter choices are 
suitable for the implementation described here at resolutions of ~100--500pc.

\section{Chemodynamics of an L$_*$ Galaxy}
\label{fiducial}

We introduce this new grid-based chemodynamical 
tool to the community.  We now demonstrate the efficacy of \textsc{Ramses-CH} 
through a presentation of the chemical properties of \texttt{109-CH}. 
Future papers in this series will extend beyond this initial demonstration into a comprehensive chemical 
tagging and chemodynamical exploration of a suite of higher-resolution cosmological 
disks spanning a range of environments and assembly histories, realisations of which are described in \citet{pilkington12a}.

\begin{figure}
\includegraphics[height=84mm]{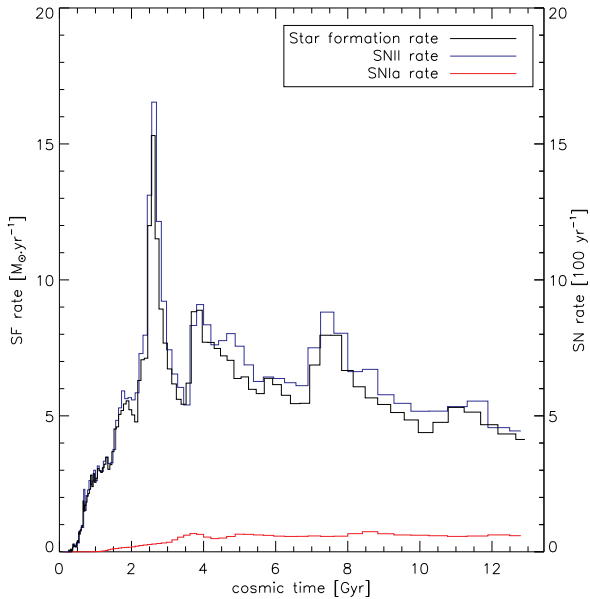}
\caption{The star formation rate for \texttt{109-CH} is shown in black 
(refer to the left-hand ordinate) and the corresponding SNeII and SNeIa 
rates are shown (refer to the right-hand ordinate) in blue and red, 
respectively. Note that the SNII rate is not precisely proportional to the star 
formation rate as would be the case for data simulated using the ``standard'' version 
of \textsc{Ramses}. The SNII rate at each time is now dependant on the star formation rate of past 
as well as present time-steps.}
\label{snrate}
\end{figure}

The first metric to consider when introducing 
SNeII and SNeIa in parallel with a relaxation of the instantaneous 
recycling approximation is the predicted supernovae rates and a 
comparison with empirical constraints.  We show the time evolution of 
these rates in Fig.~\ref{snrate}. For \texttt{109-CH}, a 
present-day SNeIa rate of 0.131~SNuM (SNe per century per 
$10^{10}$M$_\odot$ stellar mass) was found, and a SNeII rate of 0.959~SNuM. Both 
the absolute values and SNeII:SNeIa ratio ($\sim$7) are consistent with 
those found by \citet{mannucci08} for field Sbc/d galaxies: 
0.140$^{+0.045}_{-0.035}$~SNuM for SNeIa and 
0.652$^{+0.164}_{-0.134}$~SNuM for SNeII. 

\begin{figure}
\includegraphics[width=84mm]{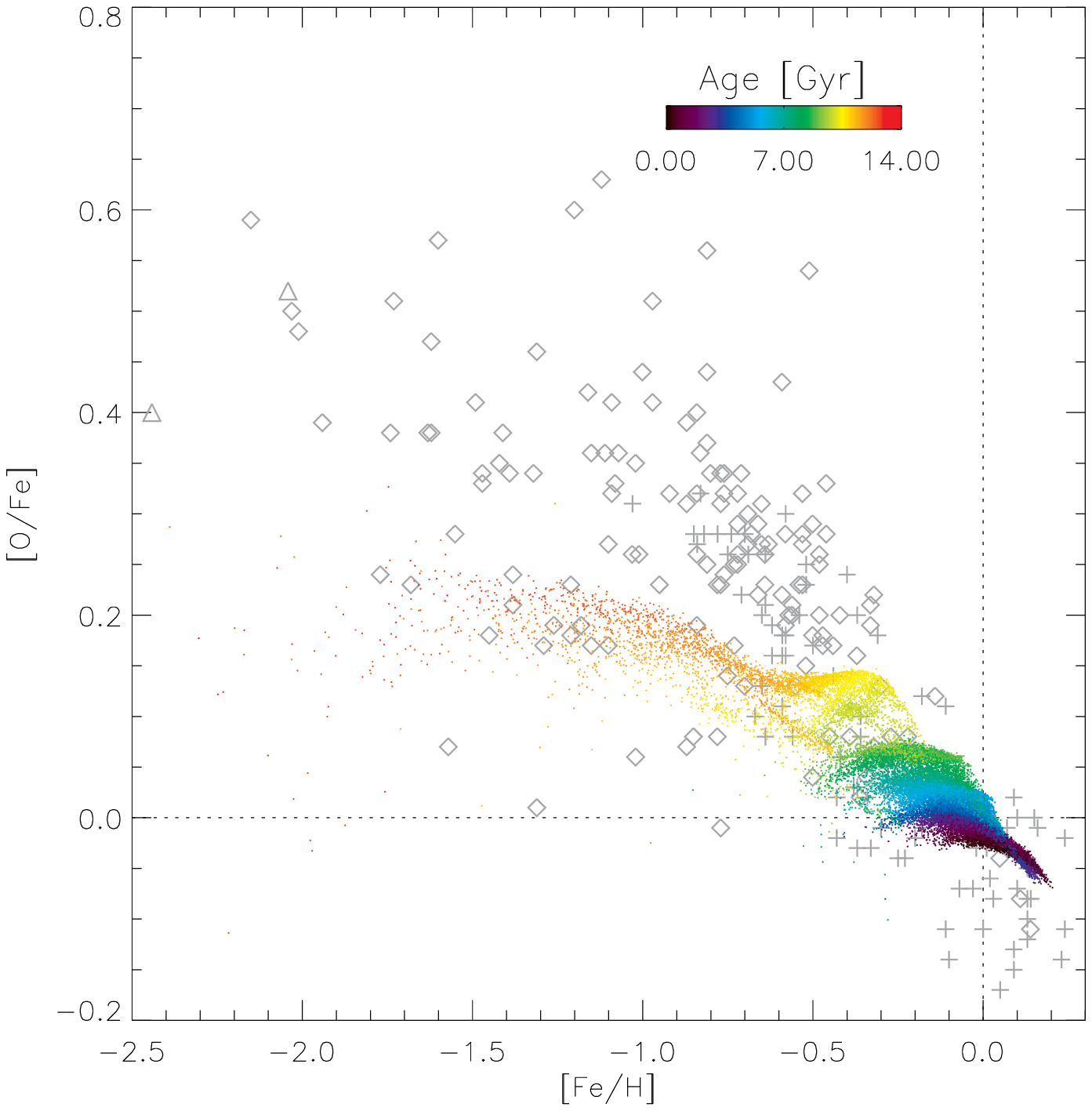}
\caption{Abundance ratios of stars in a disk region of galactocentric radius between 5 and 11 kpc
and within 3 kpc of the disk plane. Particles are 
coloured according to their age. Observational data is plotted in grey, 
triangles are very metal-poor stars from \citet{cayrel04}, diamonds are 
thick disk and halo stars from \citet{gratton03}, plusses are disk dwarf stars 
from \citet{edvardsson93}. All data has been normalised to the solar abundance 
determination of \citet{anders89}.}
\label{alphairon}
\end{figure}

Moving beyond the SNe rates, the abundance ratios of readily observed 
elements are regularly employed to constrain the timescales of star 
formation, and therefore both feedback and fundamental nucleosynthesis.  
The recovery of empirical trends found locally in the solar 
neighbourhood is a necessity for any CE model. Such observations demonstrate a clear 
correlation between $\alpha$-element and iron abundances, in the sense 
of their being an $\alpha$-enhanced plateau at lower metallicities 
(below, say, [Fe/H]$\sim$$-$0.7) with a systematic decline to solar 
values, at higher metallicities 
\citep[e.g.][]{edvardsson93,gratton03,reddy03,cayrel04,bensby05,reddy06}. This 
empirical behaviour in [$\alpha$/Fe]$-$[Fe/H] can be seen in 
Fig.~\ref{alphairon}, where, in this case, we are showing the 
observational trends for [O/Fe] vs [Fe/H].  Also shown in 
Fig.~\ref{alphairon} is the same distribution of [O/Fe]$-$[Fe/H] for the 
star particles at redshift $z$=0 within an analogous `solar 
neighbourhood' for the simulated disk \texttt{109-CH}. We should emphasise that the solar normalisation 
employed for both simulated and observed data is that of \citet{anders89}. 

A rigorous analysis of how the choice of SNeII, SNeIa, and AGB yields, 
in addition to the IMF and SNeIa progenitor model impact upon the 
chemodynamical evolution is left to future papers in this series, but it 
should be clear that even with this first `test', the qualitative 
chemical properties are not inconsistent with observations of the local 
plateau$+$decline behaviour seen in the [$\alpha$/Fe]$-$[Fe/H] 
plane.  Abundance ratios are recovered and the qualitative 
behaviour of the [Fe/H]$\sim$$-$0.7 knee is also seen in the simulated 
data. The knee-feature seen in Fig.~\ref{alphairon} at [O/Fe]=0.15, [Fe/H]=-0.2 
is attributed to the bursty star formation profile which naturally creates
multiple knee features as the SNII:SNIa rate fluctuates.

Examination of the age distribution (represented by the 
colour-coding shown in the inset to Fig.~\ref{alphairon}) reveals rapid 
early enrichment in [Fe/H], similar to the age-metallicity relations 
predicted by classical CE models. Specifically, it takes 
$\sim$3~Gyr to reach a metallicity [Fe/H]$\approx$$-$0.4, driven by the 
initial phases of intense star formation, after which the 
age-metallicity relation flattens and the rate of growth of [Fe/H] 
consequently slows (even while SNeIa are becoming more important).  This 
phase is characterised by the abundance ratio `strata' see in 
Fig.~\ref{alphairon}, with discrete `arcs' appearing at decreasing values 
of [O/Fe] as time progresses.

The influence of the cosmological environment of this simulation is 
apparent in the abundance patterns of the galaxy. The sub-sample 
displayed in Fig.~\ref{alphairon} exhibits the 
signature of merger events, e.g. the feature at [O/Fe]=0.15, [Fe/H]=-0.2.
The full extent of this is only apparent when examining all stars in the galaxy where discrete 
`streams' with chemical properties distinct from the rest of the galaxy are seen. These arise from accretion of 
satellites that have lower [Fe/H] values and remain chemically distinct. Larger 
mergers may have a similar abundance to the primary galaxy but bring gas that can 
reignite a quiescent galaxy and accelerates the production of Fe in the short term.

\section{Summary}

We present a new chemodynamical simulation code that produces feedback to account for long-lived stars and the elements 
that they produce. The more sophisticated SN feedback scheme improves the 
kinematic properties of the stellar fraction (this will be detailed in future work) and gives access to additional constraints on 
the sub-grid physics. It is clear that the galactic CE for our $L_\star$ galaxy 
does not perfectly reproduce Milky Way observations, however we believe it serves 
to demonstrate the validity of this approach. Future work using this code may merit the inclusion 
of cutting edge nucleosynthesis models \citep[e.g.][]{doherty10} and will make a comprehensive 
study of the influence of the parameters involved in the underlying CE model.

\section{Acknowledgments}

The authors thank the referee for their helpful comments. 
CGF acknowledges the support of the Science \& Technology Facilities Council (ST/F007701/1). BKG acknowledges the support of the STFC 
(ST/F002432/1; ST/G003025/1), Monash University (the Kevin Westfold Distinguished Visitor Programme)
 and Saint Mary's University. 
SC acknowledges support from the BINGO Project (ANR-08-BLAN-0316-01) and the CC-IN2P3 
Computing Center (Lyon/Villeurbanne, France), a partnership between CNRS/IN2P3 and CEA/DSM/Irfu.
DK acknowledges the support of the STFC (ST/H00260X/1). Computing resources were 
provided by the UK National Cosmology Supercomputer (COSMOS), the University of 
Central Lancashire's HPC facility and the HPC resources of CINES under the allocation 2010-c2011046642 made by GENCI (Grand Equipement National de Calcul Intensif).

\bibliographystyle{mn2e}
\bibliography{ramsesch}

\label{lastpage}
\end{document}